\documentclass{article}
\usepackage{arxiv}
\usepackage[utf8]{inputenc}
\usepackage[T1]{fontenc}
\usepackage{microtype}
\usepackage{amsmath,amssymb,amsfonts}
\usepackage{mathtools}
\usepackage{booktabs}
\usepackage{multirow}
\usepackage{graphicx}
\usepackage[dvipsnames,svgnames,x11names]{xcolor}
\usepackage{hyperref}
\usepackage{cleveref}
\usepackage{enumitem}
\usepackage{algorithm}
\usepackage{algorithmic}
\usepackage[numbers,sort&compress]{natbib}
\usepackage{caption}
\usepackage{subcaption}
\usepackage{tabularx}
\usepackage{float}
\usepackage[section]{placeins}
\usepackage{colortbl}
\usepackage{nicefrac}
\usepackage{url}

\graphicspath{ {./figures/} }
\hypersetup{
  colorlinks=true,
  linkcolor=NavyBlue,
  citecolor=NavyBlue,
  urlcolor=NavyBlue,
}

\renewcommand{\headeright}{}
\renewcommand{\undertitle}{}
\renewcommand{\shorttitle}{CUE-R: Beyond the Final Answer in RAG}

\makeatletter
\renewcommand{\paragraph}[1]{\vspace{0.8ex}\noindent\textbf{#1}\enspace}
\newcommand{\blockparagraph}[1]{\vspace{1ex}\noindent\textbf{#1}\\}
\makeatother

\newcommand{\cuer}{\textsc{Cue-R}}

\newcommand{\eg}{\textit{e.g.}}

\newcommand{\trace}{\tau}
\newcommand{\ctrace}{\tilde{\tau}}
\newcommand{\Ucorr}{U_{\text{corr}}}
\newcommand{\Ugrnd}{U_{\text{grnd}}}
\newcommand{\Ucal}{U_{\text{cal}}}

\newcommand{\Irem}{I_{\text{rem}}}
\newcommand{\Irep}{I_{\text{rep}}}
\newcommand{\Idup}{I_{\text{dup}}}
\newcommand{\finding}[1]{%
  \vspace{6pt}\noindent
  \colorbox{gray!8}{%
    \parbox{0.965\linewidth}{%
      \small\strut\textbf{Finding:}\enspace #1\strut
    }%
  }%
  \vspace{6pt}%
}

\title{CUE-R: Beyond the Final Answer in Retrieval-Augmented Generation}

\author{
  Siddharth Jain \\
  Intuit\\
  \And
  Venkat Narayan Vedam \\
  Intuit\\
}

\date{}

\begin{document}
\maketitle

\begin{abstract}
As language models shift from single-shot answer generation toward multi-step reasoning that retrieves and consumes evidence mid-inference, evaluating the role of individual retrieved items becomes more important. Existing RAG evaluation typically targets final-answer quality, citation faithfulness, or answer-level attribution, but none of these directly targets the intervention-based, per-evidence-item utility view we study here. We introduce \cuer{}, a lightweight intervention-based framework for measuring \textbf{per-evidence-item operational utility} in single-shot RAG using \textbf{shallow observable retrieval-use traces}. \cuer{} perturbs individual evidence items via \textsc{remove}, \textsc{replace}, and \textsc{duplicate} operators, then measures changes along three utility axes (correctness, proxy-based grounding faithfulness, and confidence error) plus a trace-divergence signal. We also outline an operational evidence-role taxonomy for interpreting intervention outcomes. Experiments on HotpotQA and 2WikiMultihopQA with Qwen-3 8B and GPT-5.2 reveal a consistent pattern: \textsc{remove} and \textsc{replace} substantially harm correctness and grounding while producing large trace shifts, whereas \textsc{duplicate} is often answer-redundant yet not fully behaviorally neutral. A zero-retrieval control confirms that these effects arise from degradation of meaningful retrieval. A two-support ablation further shows that multi-hop evidence items can interact non-additively: removing both supports harms performance far more than either single removal. Our results suggest that answer-only evaluation misses important evidence effects and that intervention-based utility analysis is a practical complement for RAG evaluation.
\end{abstract}

\section{Introduction}
\label{sec:intro}

Large language models have evolved from single-shot retrieve-then-answer pipelines~\citep{lewis2020retrieval,borgeaud2022improving,guu2020realm} into reasoning-oriented systems that actively fetch, revise, and verify evidence mid-generation before arriving at a conclusion~\citep{yao2023react,asai2024selfrag,jiang2023active}. In these newer architectures, retrieved evidence is not just appended context; it is consumed at intermediate reasoning steps, making its role harder to assess from the final answer alone. Yet evaluation has not kept pace with this shift. Two gaps stand out. First, final-answer quality is too coarse to explain how retrieval shapes system behavior~\citep{liu2024lost,chen2024benchmarking}. Second, free-form reasoning traces are not always reliable proxies for hidden internal computation~\citep{lanham2023measuring,turpin2023language}, which has pushed the field toward grounded evaluation using observable tool interactions, state snapshots, and intervention-based analysis~\citep{liu2024agentbench,li2023apibank}.

A gap remains. Existing work near this space typically evaluates one of three objects. Some methods evaluate \emph{final-answer quality}, asking only whether the system answered correctly~\citep{yang2018hotpotqa,ho2020constructing}. Others evaluate \emph{trace faithfulness}, asking whether a visible reasoning process appears grounded or valid~\citep{creswell2023selection,wei2022chain}. Still others evaluate \emph{answer-level attribution}, asking whether removing a retrieved evidence item changes the final answer~\citep{gao2023enabling,bohnet2022attributed}. Each of these perspectives is useful, but none fully answers the question we target:

\begin{quote}
\emph{What role did a retrieved evidence item play once the system acted on it?}
\end{quote}

\noindent The value of this question is not merely to show that harmful perturbations can hurt performance (which is expected) but to characterize evidence effects that are not well summarized by final-answer accuracy alone.

This question matters because retrieved evidence can affect model behavior in ways not cleanly captured by answer correctness. An evidence item may be indispensable for correctness, useful only for grounding, redundant with respect to the final answer, harmful because it diverts the system toward an unstable path, or confidence-distorting because it changes certainty without improving performance. Recent work on misleading or conflicting retrieval~\citep{xie2024adaptive,pan2023risk}, process-faithful reasoning~\citep{creswell2023selection}, and attribution sensitivity~\citep{gao2023enabling} shows that these distinctions matter in practice for evaluating modern retrieval-augmented systems.

We introduce \cuer{}, a lightweight framework for measuring the \textbf{operational intervention-based utility of retrieved evidence} over \textbf{shallow observable retrieval-use traces}. \cuer{} treats retrieval evaluation as an intervention problem. Given a question, a retrieved evidence set, and an observable trace, we perturb one evidence item at a time using three operators (\textsc{remove}, \textsc{replace}, and \textsc{duplicate}) and then measure how the perturbation changes both the observable trace and downstream utility along three axes: correctness, proxy-based grounding faithfulness, and confidence error. In contrast to answer-only evidence ablations, \cuer{} is designed to capture cases where a perturbation leaves the final answer unchanged while still altering the model's behavior in important ways.

The central empirical motivation for \cuer{} is simple. In our experiments, harmful perturbations frequently induce large trace changes even when answer-level metrics remain unchanged. For example, on HotpotQA with Qwen-3 8B, both \textsc{remove} and \textsc{replace} interventions yield large mean trace divergence ($\approx$0.61--0.63), while \textsc{duplicate} interventions are much milder on correctness but still measurably affect confidence error. A zero-retrieval control further shows that retrieval is genuinely useful in the base setting, so the intervention effects reflect degradation of meaningful evidence rather than arbitrary prompt instability. This pattern replicates across 2WikiMultihopQA and a second model family (GPT-5.2), suggesting that the phenomenon is not confined to a single dataset or model. A two-support ablation further shows that evidence items can interact non-additively in multi-hop questions, with joint removal causing degradation that exceeds either single removal.

\cuer{} is deliberately limited in scope. It does not attempt to recover hidden internal reasoning or provide a mechanistic theory of model cognition. It focuses on \textbf{shallow observable traces}: logged evidence identifiers, candidate answers, support status, and confidence proxies. In the current instantiation, these traces are limited to a single-shot RAG pipeline; extending the approach to deeper agentic workflows is future work. Given current uncertainty about the faithfulness of free-form reasoning traces~\citep{lanham2023measuring}, observable and auditable system behavior is a reasonable starting point for an intervention-based evidence framework.

\blockparagraph{Contributions.}
\begin{enumerate}[leftmargin=*,itemsep=2pt]
  \item We introduce a \textbf{lightweight intervention-based framework} for measuring per-evidence-item utility in single-shot RAG using shallow observable retrieval-use traces.
  \item We propose a \textbf{multi-axis evaluation view} that separates correctness, proxy grounding, confidence error, and trace divergence, enabling analysis beyond final-answer accuracy.
  \item Across two datasets and two model families, we show that \textsc{remove} and \textsc{replace} are \textbf{consistently harmful}, while \textsc{duplicate} is often answer-redundant yet not fully behaviorally neutral.
  \item A two-support ablation shows that evidence items can exhibit \textbf{non-additive interaction}: removing both supports causes far larger degradation than either single removal, revealing joint dependence that single-item interventions miss.
\end{enumerate}

\noindent We also provide an operational evidence-role taxonomy as an interpretive lens over these intervention outcomes.

\section{Related Work}
\label{sec:related}

\subsection{Answer-Level Attribution in RAG}
\label{sec:rw-attribution}

Prior work on attribution of retrieved evidence to final answers is the closest antecedent to \cuer{}. \citet{petroni2021kilt} established a benchmark for knowledge-intensive language tasks, while \citet{gao2023enabling} and \citet{bohnet2022attributed} argued that similarity-based attribution is often only plausibly explanatory rather than causal, and proposed evaluating citation quality and answer-level attribution in generated text.

\cuer{} builds on this line of work but differs in two ways. First, our target object is not only the final answer but the \emph{observable retrieval-use trace} followed by the system. Second, we do not treat attribution as a scalar contribution score alone. Instead, we decompose evidence effects across correctness, grounding faithfulness, and confidence alignment, and we explicitly distinguish constructive, redundant, distractive, and confidence-distorting evidence roles.

\subsection{Faithfulness in Retrieval-Augmented Reasoning}
\label{sec:rw-faithfulness}

A second nearby literature studies whether the reasoning process itself is faithful to available evidence. \citet{asai2024selfrag} propose Self-RAG, which trains models to retrieve, generate, and self-critique with reflection tokens. \citet{creswell2023selection} argue that evaluation should reward intermediate traceability and faithful reasoning. \citet{lanham2023measuring} and \citet{turpin2023language} raise concerns about whether chain-of-thought traces faithfully reflect the model's internal computation.

\cuer{} shares this motivation but asks a narrower question: \emph{what utility did a specific retrieved evidence item contribute once the system acted on it?} The difference is practical: a perturbation can leave correctness unchanged while still altering the trace or confidence alignment.

\subsection{Grounded Evaluation from Observable Traces}
\label{sec:rw-traces}

Recent work on agent evaluation has moved toward grounded assessment using observable tool traces and system state. \citet{liu2024agentbench} introduce a benchmark for evaluating LLMs as agents across diverse environments. \citet{li2023apibank} propose API-Bank for evaluating tool-augmented LLMs. \citet{yao2023react} demonstrate that interleaving reasoning and acting through observable action traces improves task performance and interpretability.

\cuer{} shares this evaluation philosophy, treating observable behavior as the primary object of analysis. However, our problem setting is different: existing grounded trace benchmarks evaluate trajectories broadly, while \cuer{} focuses specifically on retrieved evidence items and their counterfactual role in retrieval-augmented reasoning.

\subsection{Misleading, Noisy, and Conflicting Retrieval}
\label{sec:rw-noisy}

Another closely related line of work studies robustness of RAG systems under misleading or conflicting evidence. \citet{liu2024lost} show that language models struggle when relevant information appears in the middle of long contexts. \citet{chen2024benchmarking} benchmark RAG under various noise and conflicting conditions. \citet{pan2023risk} and \citet{xie2024adaptive} study risks from factual conflicts between parametric and retrieved knowledge.

Our \textsc{replace} and \textsc{duplicate} interventions approximate harmful evidence effects emphasized by this robustness literature. \cuer{} contributes a different layer of analysis: rather than only reporting whether performance falls under conflict or noise, we ask \emph{how} a perturbed evidence item changes trace behavior and utility, enabling us to distinguish between strongly distractive, answer-redundant, and confidence-distorting evidence.

\subsection{Recent RAG Evaluation and Counterfactual Attribution}
\label{sec:rw-recent}

The RAG evaluation landscape has expanded rapidly. \citet{yu2024evaluation} survey the growing literature on retrieval-augmented generation evaluation, identifying evaluation as a key open challenge beyond simple answer quality. \citet{ru2024ragchecker} introduce RAGChecker, a fine-grained diagnostic framework that decomposes RAG failures into retriever-side and generator-side issues using claim-level entailment. \citet{wallat2025correctness} distinguish citation correctness (does the document support the statement?) from citation faithfulness (did the model genuinely rely on the document?), finding that up to 57\% of citations lack faithfulness. \citet{stolfo2024groundedness} empirically measure groundedness in retrieval-augmented long-form generation, finding significant ungroundedness even in correct answers. Most directly related to our work, \citet{saharoy2025evidence} use counterfactual evidence removal to measure evidence contribution in conversational QA with RAG systems.

The field has moved past answer-only evaluation, so it is worth clarifying how \cuer{} differs. \cuer{} is \emph{not} an answer-level attribution method, a citation quality evaluator, a faithfulness classifier, a hallucination detector, or a trajectory evaluator for general-purpose agents. It is specifically an \textbf{intervention-based framework} for diagnosing the operational utility of individual retrieved evidence items in single-shot RAG through controlled perturbation and multi-axis outcome measurement. Compared to \citet{saharoy2025evidence}, who focus on answer-level attribution via removal, \cuer{} additionally measures grounding, confidence error, and trace divergence under three distinct perturbation types (not only removal). Compared to \citet{ru2024ragchecker}, who diagnose RAG at the claim level, \cuer{} targets per-evidence-item utility through intervention rather than post-hoc entailment checking.

\subsection{Positioning of CUE-R}
\label{sec:rw-positioning}

\Cref{tab:positioning} summarizes how \cuer{} relates to neighboring evaluation perspectives.

\begin{table}[!htbp]
\centering
\caption{Positioning of \cuer{} relative to neighboring evaluation perspectives.}
\label{tab:positioning}
\small
\begin{tabular}{@{}lllc@{}}
\toprule
\textbf{Perspective} & \textbf{Target Object} & \textbf{Intervention?} & \textbf{Multi-Axis?} \\
\midrule
Answer-level attribution & Final answer & No & No \\
Reasoning-trace faithfulness & Reasoning chain & No & No \\
Robustness benchmarking & System accuracy & Implicit & No \\
Trajectory evaluation & Full agent trace & No & No \\
Fine-grained RAG diagnostics & Claim-level entailment & No & Yes \\
Counterfactual attribution & Answer change under removal & Yes & No \\
\midrule
\textbf{\cuer{} (ours)} & \textbf{Evidence utility over trace} & \textbf{Yes} & \textbf{Yes} \\
\bottomrule
\end{tabular}
\end{table}

\cuer{} sits between these views. It is not a hidden-mechanism explanation method, not a generic trace-faithfulness benchmark, and not only an answer-level attribution score. It is a \textbf{diagnostic intervention framework} that reveals both marginal evidence necessity and multi-item interaction effects over shallow observable retrieval-use traces.

\section{Problem Formulation}
\label{sec:formulation}

\subsection{Setup}
\label{sec:setup}

We consider a retrieval-augmented system that receives a question $q$, retrieves a set of evidence candidates from a corpus $\mathcal{C}$, and produces a final answer $y$. Unlike answer-only formulations, we assume the system exposes an \emph{observable retrieval-use trace} that records coarse actions and state snapshots during inference. In the present paper, this general formulation is instantiated only in a shallow single-shot setting, where the observable trace reduces to retrieved context, model-reported used chunk identifiers, answer, confidence, and brief rationale.

Formally, a run is represented as a trace:
\begin{equation}
\label{eq:trace}
\trace = \bigl((s_0, a_1),\, (s_1, a_2),\, \ldots,\, (s_{T-1}, a_T),\, s_T,\, y,\, c\bigr),
\end{equation}
where $a_t$ is the action at step $t$, $s_t$ is the observable state after that step, $y$ is the final answer, and $c \in [0,1]$ is a confidence signal or calibrated proxy.

The action space is standardized to a small vocabulary:
\begin{equation}
\label{eq:actions}
a_t \in \{\textsc{retrieve},\; \textsc{select},\; \textsc{verify},\; \textsc{infer},\; \textsc{answer}\}.
\end{equation}

Each observable state is defined as:
\begin{equation}
\label{eq:state}
s_t = (q_t,\, E_t,\, v_t,\, \hat{y}_t,\, c_t),
\end{equation}
where $q_t$ is the active query or reformulated sub-query, $E_t \subseteq \mathcal{E}$ is the set of selected evidence identifiers, $v_t$ is verification or support status, $\hat{y}_t$ is the candidate answer at step $t$, and $c_t$ is a step-level confidence proxy.

This formulation deliberately avoids stronger claims about hidden internal reasoning. The object of study is the observable behavior of the system under evidence perturbation.

\subsection{Evidence Intervention Problem}
\label{sec:intervention}

Let $R(q, \mathcal{C}) = \{e_1, \ldots, e_k\}$ denote the retrieved evidence set for question $q$. Standard evaluation asks whether the final answer $y$ is correct. \cuer{} asks a different question:

\begin{quote}
\emph{For a retrieved evidence item $e_i$, what role did it play in producing the observed trace and final outcome?}
\end{quote}

To answer this, we define an intervention operator $I$ applied to an evidence item $e_i$, producing a counterfactual trace:
\begin{equation}
\label{eq:ctrace}
\ctrace_{e_i}^{\,I} \;.
\end{equation}

In this paper, we consider three intervention types:
\begin{equation}
\label{eq:interventions}
I \in \{\Irem,\; \Irep,\; \Idup\},
\end{equation}
corresponding to removing the target evidence item, replacing it with a non-supporting alternative, and duplicating it in the evidence set.

\subsection{Multi-Axis Utility}
\label{sec:utility}

A single scalar notion of performance is not sufficient for this comparison. A perturbation may leave the final answer unchanged while worsening grounding or confidence alignment, or it may preserve correctness while substantially changing the path taken by the system.

We define utility along three primary axes:
\begin{align}
\label{eq:utility}
\Ucorr(\trace) &\quad \text{(answer correctness)}, \nonumber \\
\Ugrnd(\trace) &\quad \text{(grounding faithfulness, proxy-based)}, \\
\Ucal(\trace)  &\quad \text{(confidence error)}. \nonumber
\end{align}

\noindent Trace divergence (\Cref{sec:divergence}) is treated as a separate behavioral signal rather than a fourth utility axis, because it captures path-level change rather than outcome-level quality.

For intervention $I$ on evidence item $e_i$, the intervention-induced utility change is:
\begin{equation}
\label{eq:delta}
\Delta_{e_i}^{(k)}(I) = U_k(\trace) - U_k\bigl(\ctrace_{e_i}^{\,I}\bigr), \quad k \in \{\text{corr},\, \text{grnd},\, \text{cal}\}.
\end{equation}

This definition captures the counterfactual contribution of the evidence item along each axis.

\subsection{Trace Divergence}
\label{sec:divergence}

Utility deltas describe outcome change but not necessarily behavioral change. Two traces may end with the same answer while taking very different paths. To capture this, we define a trace-divergence function:
\begin{equation}
\label{eq:div-general}
D(\trace,\, \ctrace_{e_i}^{\,I}) = \alpha_1\, d_{\text{act}}(\trace, \ctrace') + \alpha_2\, d_{E}(\trace, \ctrace') + \alpha_3\, d_{\text{ans}}(\trace, \ctrace') + \alpha_4\, d_{v}(\trace, \ctrace'),
\end{equation}
with $\alpha_i \geq 0$ and $\sum_i \alpha_i = 1$, combining action-sequence edit distance, evidence-set Jaccard divergence, candidate-answer divergence, and verification-status disagreement.

For the shallow single-shot setting in our experiments, we use a simplified proxy:
\begin{equation}
\label{eq:div-proxy}
D(\trace, \trace') = 0.5 \cdot d_{\text{Jaccard}}(E, E') + 0.3 \cdot \mathbb{1}[y \neq y'] + 0.2 \cdot |c - c'|,
\end{equation}
where $d_{\text{Jaccard}}(E, E') = 1 - \frac{|E \cap E'|}{|E \cup E'|}$ is the Jaccard divergence over used evidence identifiers, $\mathbb{1}[y \neq y']$ is the answer-change indicator, and $|c - c'|$ is the absolute confidence change.

\subsection{Evidence-Role Taxonomy}
\label{sec:taxonomy}

Given the utility deltas and trace divergence, \cuer{} uses intervention outcomes to suggest an operational role for each evidence item:

\begin{itemize}[leftmargin=*,itemsep=2pt]
  \item \textbf{Constructive:} perturbing the item harms correctness or grounding ($\Delta^{(\text{corr})} > 0$ or $\Delta^{(\text{grnd})} > 0$).
  \item \textbf{Corrective:} perturbing the item removes a recovery path from an otherwise wrong or unsupported intermediate state. (This role is the hardest to operationalize from shallow traces alone; reliably detecting it requires deeper multi-step traces where intermediate recovery is observable.)
  \item \textbf{Redundant:} perturbing causes negligible utility change \emph{and} negligible trace divergence ($|\Delta^{(k)}| \approx 0$, $D \approx 0$).
  \item \textbf{Distractive:} perturbation induces large trace change and worsens downstream utility ($D \gg 0$, $\Delta^{(k)} > 0$ for correctness/grounding, indicating the original outperforms the perturbed run).
  \item \textbf{Confidence-distorting:} perturbation changes confidence error without corresponding correctness benefit ($|\Delta^{(\text{cal})}| > 0$, $\Delta^{(\text{corr})} \approx 0$).
\end{itemize}

This role assignment is an operational label derived from observable counterfactual behavior, not a metaphysical statement about internal causality.

\subsection{Framework Objective}
\label{sec:objective}

The framework objective is to diagnose retrieved evidence through intervention. Given a question $q$, retrieved evidence set $R(q, \mathcal{C})$, and observed trace $\trace$, \cuer{} outputs for each intervened item $e_i$:
\begin{equation}
\label{eq:output}
\Bigl(\Delta_{e_i}^{(\text{corr})},\; \Delta_{e_i}^{(\text{grnd})},\; \Delta_{e_i}^{(\text{cal})},\; D(\trace, \ctrace_{e_i}^{\,I}),\; \textsc{role}(e_i)\Bigr).
\end{equation}

\section{Method}
\label{sec:method}

\subsection{Overview}
\label{sec:method-overview}

\begin{figure}[!htbp]
\centering
\includegraphics[width=\linewidth]{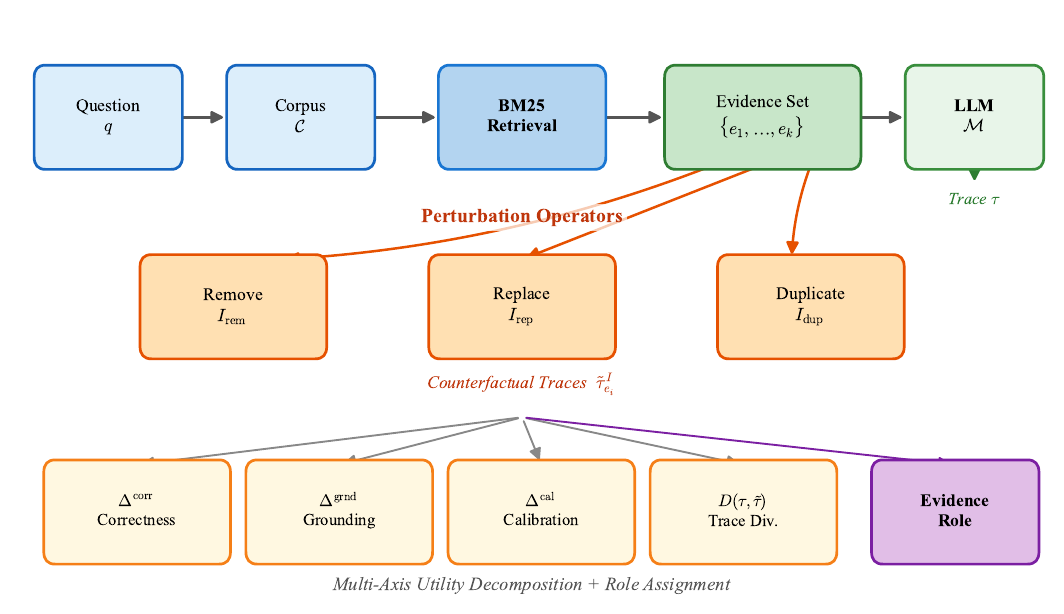}
\caption{\textbf{CUE-R framework overview.} A question $q$ and corpus $\mathcal{C}$ are passed through BM25 retrieval to produce an evidence set and original trace $\tau$. Three perturbation operators (\textsc{remove}, \textsc{replace}, \textsc{duplicate}) are applied to a target evidence item, producing counterfactual traces. Multi-axis utility decomposition and role assignment diagnose the evidence item's contribution.}
\label{fig:framework}
\end{figure}

\cuer{} is an intervention-based evaluation framework for retrieval-augmented reasoning systems. Given a question, a retrieved evidence set, and an observable retrieval-use trace, \cuer{} measures how perturbing retrieved evidence changes both the observable trace and the downstream utility of the run. The framework is designed for settings where systems expose externally observable behavior such as retrieval calls, selected evidence, candidate answers, verification status, and final outputs. \cuer{} does not attempt to recover hidden internal reasoning; its target is strictly the interventional sensitivity of observable traces under evidence perturbation. \Cref{fig:framework} provides an overview of the pipeline.

\subsection{Standardized Observable Trace}
\label{sec:method-trace}

A run is represented as a standardized observable trace (\Cref{eq:trace}), with actions restricted to a coarse vocabulary (\Cref{eq:actions}) and states defined as tuples (\Cref{eq:state}). In the single-shot RAG setting used in our experiments, the trace collapses to retrieval, evidence selection, and answer generation. The same schema naturally extends to agentic settings with additional retrieve--select--verify loops.

\subsection{Evidence Perturbations}
\label{sec:method-perturb}

Let $e$ denote a retrieved evidence item selected for intervention. \cuer{} applies three perturbation operators (\Cref{eq:interventions}):
\begin{itemize}[leftmargin=*,itemsep=2pt]
  \item \textbf{Remove} ($\Irem$): Delete the target item from the context entirely.
  \item \textbf{Replace} ($\Irep$): Substitute the target with a topically related but non-supporting passage.
  \item \textbf{Duplicate} ($\Idup$): Add a second copy of the target to the evidence set.
\end{itemize}
These operators probe different failure modes: \textsc{remove} tests evidence necessity, \textsc{replace} tests robustness to misleading substitution, and \textsc{duplicate} tests sensitivity to redundancy.

\subsection{Utility Decomposition and Trace Divergence}
\label{sec:method-utility}

Utility is measured along correctness, grounding, and confidence-error axes (\Cref{eq:utility,eq:delta}). Trace divergence (\Cref{eq:div-general,eq:div-proxy}) captures behavioral changes that utility deltas alone may miss: two traces can produce the same answer while following very different paths.

\subsection{Evidence-Role Taxonomy}
\label{sec:method-taxonomy}

Evidence items can be interpreted through an operational taxonomy of constructive, corrective, redundant, distractive, and confidence-distorting roles based on observed utility changes under intervention (\Cref{sec:taxonomy}). This interpretation is operational, not mechanistic.

\subsection{Practical Instantiation}
\label{sec:method-practical}

Our experiments use a single-shot RAG pipeline over HotpotQA~\citep{yang2018hotpotqa} and 2WikiMultihopQA~\citep{ho2020constructing}. For each question:
\begin{enumerate}[leftmargin=*,itemsep=2pt]
  \item Retrieve the top $k = 5$ chunks using BM25~\citep{robertson2009probabilistic}.
  \item Prompt the model to answer using only the provided context.
  \item Log selected chunk identifiers, answer, confidence, and brief rationale.
  \item Rerun under \textsc{original}, \textsc{remove}, \textsc{replace}, and \textsc{duplicate} settings.
\end{enumerate}

Answer correctness is computed using both strict (normalized exact match) and soft matching (with yes/no canonicalization, number normalization, and high-overlap fuzzy matching at F1 $\geq$ 0.8). Grounding faithfulness is approximated by overlap between model-used chunk identifiers and gold supporting titles. Confidence error is measured as the absolute difference between the model's self-reported confidence and binary correctness.

\section{Experimental Setup}
\label{sec:experiments}

\subsection{Evaluation Goals}
\label{sec:goals}

Our experiments address five questions:
\begin{enumerate}[label=\textbf{Q\arabic*.},leftmargin=2.5em,itemsep=2pt]
  \item Do different evidence perturbations induce distinct utility profiles?
  \item Does trace-sensitive evaluation reveal effects that answer-only metrics miss?
  \item Do these patterns replicate across datasets?
  \item Do they replicate across model families?
  \item Can \cuer{} reveal non-additive interaction effects between evidence items in multi-hop questions?
\end{enumerate}

\subsection{Datasets}
\label{sec:datasets}

\paragraph{HotpotQA.} We use the distractor setting of HotpotQA~\citep{yang2018hotpotqa}. Each example includes a question, an answer, supporting facts, and a fixed set of candidate paragraphs. We use 200 examples for the primary Qwen-3 8B analysis, 200 for the zero-retrieval control, and 100 for the cross-model replication with GPT-5.2.

\paragraph{2WikiMultihopQA.} We use 100 examples from 2WikiMultihopQA~\citep{ho2020constructing} as a second multi-hop QA benchmark for cross-dataset replication.

\subsection{Models}
\label{sec:models}

\paragraph{Qwen-3 8B.} Our primary experiments use a local Qwen-3 8B model~\citep{qwen2025qwen3} served through Ollama (temperature = 0).

\paragraph{GPT-5.2.} We use GPT-5.2 for a lightweight cross-family replication on HotpotQA (temperature = 0).

\subsection{Retrieval Pipeline}
\label{sec:retrieval}

All experiments use the same minimal retrieval setup. Candidate paragraphs are converted into passage chunks with stable chunk identifiers (formatted as \texttt{ctx\_0}, \texttt{ctx\_1}, \ldots). BM25~\citep{robertson2009probabilistic} ranks candidate chunks against the question, and the top $k = 5$ chunks are provided to the model.

\subsection{Interventions and Target Selection}
\label{sec:target}

For each example, we first run the original retrieval condition. We then select a target evidence item using a priority heuristic:
\begin{enumerate}[leftmargin=*,itemsep=1pt]
  \item Prefer a chunk explicitly referenced by the model whose title matches a gold support title.
  \item Otherwise prefer any model-used chunk.
  \item Otherwise prefer the first retrieved chunk matching a support title.
  \item Otherwise fall back to the highest-ranked retrieved chunk.
\end{enumerate}
Three perturbations are then applied: \textsc{remove}, \textsc{replace}, and \textsc{duplicate}.

\subsection{Replacement-Hardness Sweep}
\label{sec:sweep}

To test whether the \textsc{replace} intervention is sensitive to corruption type, we run a replacement-hardness sweep on HotpotQA with Qwen-3 8B. We define three replacement modes:
\begin{itemize}[leftmargin=*,itemsep=1pt]
  \item \textbf{Easy:} a random non-support chunk.
  \item \textbf{Medium:} a question-similar non-support chunk (highest BM25 score against the question).
  \item \textbf{Hard:} a chunk most similar to the target evidence (highest BM25 score against the target text) while still being non-supporting.
\end{itemize}

\subsection{Zero-Retrieval Control}
\label{sec:zero}

To verify that retrieval is genuinely useful in the base setting, we run a zero-retrieval control on HotpotQA. In this condition, the model receives the question without any retrieved context and must answer from parametric knowledge alone.

\subsection{Metrics}
\label{sec:metrics}

We evaluate each run using five primary metrics:

\begin{itemize}[leftmargin=*,itemsep=2pt]
  \item \textbf{Soft Correctness.} Normalized match with yes/no canonicalization, number normalization, and high-overlap fuzzy matching (F1 $\geq$ 0.8).
  \item \textbf{Answer F1.} Token-level F1 between the predicted answer and gold answer.
  \item \textbf{Grounding Score (proxy).} Fraction of model-used chunk identifiers whose titles match gold supporting facts:
  \begin{equation}
  \label{eq:grounding}
  G = \frac{|\{u \in U : \textsc{title}(u) \in S\}|}{|U|},
  \end{equation}
  where $U$ is the set of used chunk IDs and $S$ is the set of gold support titles (0 if $U = \emptyset$). This is a coarse proxy: title-level matching does not verify that the model used the correct information within a chunk, and gold support titles may not exhaustively enumerate all useful evidence.
  \item \textbf{Confidence Error.} $\text{CE} = |c - \mathbb{1}[\text{is\_correct}]|$, where $c$ is the model's self-reported confidence. This is an instance-level proxy for calibration mismatch, not a distributional calibration metric in the classical sense~\citep{guo2017calibration}.
  \item \textbf{Trace Divergence.} Computed via \Cref{eq:div-proxy}.
\end{itemize}

\subsection{Statistical Analysis}
\label{sec:stats}

For the main experiments, we compute bootstrap confidence intervals (5{,}000 resamples, 95\% CI, seed $= 42$) for intervention means, and paired bootstrap deltas with two-sided $p$-values relative to the original condition. The two-sided $p$-value is computed as $p = 2 \cdot \min\bigl(P(\bar{\delta}^* \geq 0),\, P(\bar{\delta}^* \leq 0)\bigr)$, where $\bar{\delta}^*$ denotes the bootstrap distribution of the mean paired difference.

\section{Results}
\label{sec:results}

We evaluate \cuer{} across two multi-hop QA datasets and two model families. The goal is not to maximize task accuracy but to provide evidence that evidence perturbations induce distinct utility profiles not fully visible to answer-only evaluation.

\subsection{Evidence Perturbations Induce Distinct Utility Profiles (Q1)}
\label{sec:main-result}

The main result is not simply that harmful perturbations reduce answer quality, but that the intervention types produce \emph{distinct multi-axis utility profiles}. In particular, \textsc{duplicate} often preserves answer correctness while still inducing measurable shifts in trace behavior and, in some cases, grounding or confidence error, whereas \textsc{remove} and \textsc{replace} strongly degrade both answer quality and evidence use. We first evaluate on 200 HotpotQA examples using Qwen-3 8B (\Cref{tab:main}).

\begin{table}[!htbp]
\centering
\caption{\cuer{} results on HotpotQA with Qwen-3 8B ($n = 200$). The three interventions produce distinct utility profiles. Best non-original values are \textbf{bolded}.}
\label{tab:main}
\small
\begin{tabular}{@{}lccccc@{}}
\toprule
\textbf{Intervention} & \textbf{Correct.}$\uparrow$ & \textbf{Ans. F1}$\uparrow$ & \textbf{Ground.}$\uparrow$ & \textbf{Conf. Err.}$\downarrow$ & \textbf{Trace Div.}$\downarrow$ \\
\midrule
Original  & 0.585 & 0.640 & 0.823 & 0.422 & 0.000 \\
\midrule
Remove    & 0.285 & 0.329 & 0.392 & 0.639 & 0.632 \\
Replace   & 0.270 & 0.318 & 0.353 & 0.667 & 0.637 \\
Duplicate & \textbf{0.585} & \textbf{0.639} & \textbf{0.845} & \textbf{0.424} & \textbf{0.074} \\
\bottomrule
\end{tabular}
\end{table}

Relative to the original setting, both \textsc{remove} and \textsc{replace} substantially reduce answer quality, grounding, and confidence alignment, while \textsc{duplicate} is much milder. Under the original condition, mean correctness is 0.585 and mean answer F1 is 0.640. Removing the target chunk reduces correctness to 0.285 and F1 to 0.329, while replacing it reduces correctness further to 0.270 and F1 to 0.318. In contrast, duplicating the target chunk fully preserves answer behavior (correctness 0.585, F1 0.639).

\begin{figure}[!htbp]
\centering
\includegraphics[width=\linewidth]{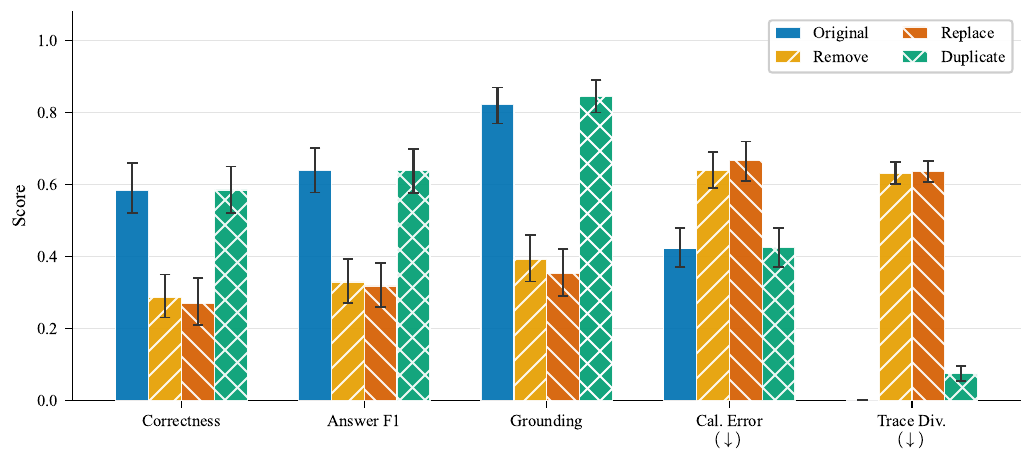}
\caption{\textbf{\cuer{} results on HotpotQA with Qwen-3 8B ($n=200$).} Grouped bar chart with 95\% bootstrap confidence intervals. \textsc{Remove} and \textsc{replace} substantially reduce correctness, F1, and grounding while increasing confidence error and trace divergence. \textsc{Duplicate} is much milder across all axes.}
\label{fig:main-results}
\end{figure}

The same asymmetry appears in grounding and confidence error. Original grounding is 0.823, but drops to 0.392 under removal and 0.353 under replacement. Confidence error worsens from 0.422 to 0.639 (remove) and 0.667 (replace). Duplication remains close to the baseline across all axes. \Cref{fig:main-results} visualizes these results with bootstrap confidence intervals.

\finding{Different evidence perturbations do not behave like generic ``noise.'' Instead, they produce qualitatively different utility profiles.}

\FloatBarrier
\subsection{Trace-Sensitive Evaluation Reveals Additional Effects (Q2)}
\label{sec:trace-result}

A central motivation for \cuer{} is that answer-only metrics alone do not cleanly separate all evidence effects. For \textsc{remove} and \textsc{replace}, trace divergence (0.632 and 0.637) is obviously coupled with answer deterioration, so trace analysis adds less independent information in those cases. The cleaner test is \textsc{duplicate}, where correctness is unchanged at 0.585 yet trace divergence is nonzero (0.074), and paired analysis shows significant trace ($p < 0.001$) and grounding ($p = 0.039$) effects despite null correctness and F1 deltas.

At the example level, many duplicate cases exhibit measurable trace divergence while preserving the same correctness outcome. These cases show that an answer-only evaluation would mark the intervention as fully benign, even though the system's evidence-use pattern changed (see qualitative examples in \Cref{sec:cases}). The trace signal is most informative precisely in these answer-preserving cases, where it reveals behavioral shifts that correctness alone cannot detect.

\subsection{Statistical Support for Intervention Effects}
\label{sec:stats-result}

\Cref{tab:deltas} presents bootstrap confidence intervals and paired deltas for all intervention comparisons.

\begin{table}[!htbp]
\centering
\caption{Paired bootstrap deltas (original $-$ intervention) for Qwen-3 8B on HotpotQA ($n = 200$; 5{,}000 resamples). Positive $\Delta$ indicates the original outperforms the intervention. $^\dagger$Significant at $p < 0.05$.}
\label{tab:deltas}
\small
\begin{tabular}{@{}llccc@{}}
\toprule
\textbf{Comparison} & \textbf{Metric} & $\boldsymbol{\Delta}$ & \textbf{95\% CI} & $\boldsymbol{p}$ \\
\midrule
\multirow{5}{*}{Orig vs Remove}
  & Correctness & $+$0.300 & [0.225, 0.375] & $<$0.001$^\dagger$ \\
  & Answer F1   & $+$0.311 & [0.237, 0.385] & $<$0.001$^\dagger$ \\
  & Grounding   & $+$0.430 & [0.350, 0.506] & $<$0.001$^\dagger$ \\
  & Conf.\ Error & $-$0.218 & [$-$0.282, $-$0.156] & $<$0.001$^\dagger$ \\
  & Trace Div.  & $-$0.632 & [$-$0.662, $-$0.601] & $<$0.001$^\dagger$ \\
\midrule
\multirow{5}{*}{Orig vs Replace}
  & Correctness & $+$0.315 & [0.240, 0.390] & $<$0.001$^\dagger$ \\
  & Answer F1   & $+$0.323 & [0.249, 0.394] & $<$0.001$^\dagger$ \\
  & Grounding   & $+$0.470 & [0.392, 0.543] & $<$0.001$^\dagger$ \\
  & Conf.\ Error & $-$0.246 & [$-$0.308, $-$0.186] & $<$0.001$^\dagger$ \\
  & Trace Div.  & $-$0.637 & [$-$0.666, $-$0.607] & $<$0.001$^\dagger$ \\
\midrule
\multirow{5}{*}{Orig vs Duplicate}
  & Correctness & 0.000 & [$-$0.025, 0.025] & 1.000 \\
  & Answer F1   & $+$0.002 & [$-$0.015, 0.017] & 0.827 \\
  & Grounding   & $-$0.023 & [$-$0.045, $-$0.001] & 0.039$^\dagger$ \\
  & Conf.\ Error & $-$0.003 & [$-$0.024, 0.019] & 0.815 \\
  & Trace Div.  & $-$0.074 & [$-$0.096, $-$0.054] & $<$0.001$^\dagger$ \\
\bottomrule
\end{tabular}
\end{table}

The original-minus-remove correctness delta is $+$0.300 [0.225, 0.375] ($p < 0.001$), while the original-minus-replace delta is $+$0.315 [0.240, 0.390] ($p < 0.001$). Grounding drops are similarly large: $+$0.430 for remove and $+$0.470 for replace (both $p < 0.001$). Confidence error also worsens significantly under both harmful interventions ($p < 0.001$). The duplicate intervention behaves differently: the correctness delta is exactly 0.000 ($p = 1.0$), and the F1 delta is only $+$0.002. However, duplication still produces a statistically significant grounding shift ($-$0.023, $p = 0.039$) and trace divergence ($-$0.074, $p < 0.001$).

\finding{Duplicate evidence is answer-redundant without being fully behaviorally neutral; it significantly alters grounding and trace behavior even when correctness is perfectly preserved.}

\begin{figure}[!htbp]
\centering
\includegraphics[width=0.95\linewidth]{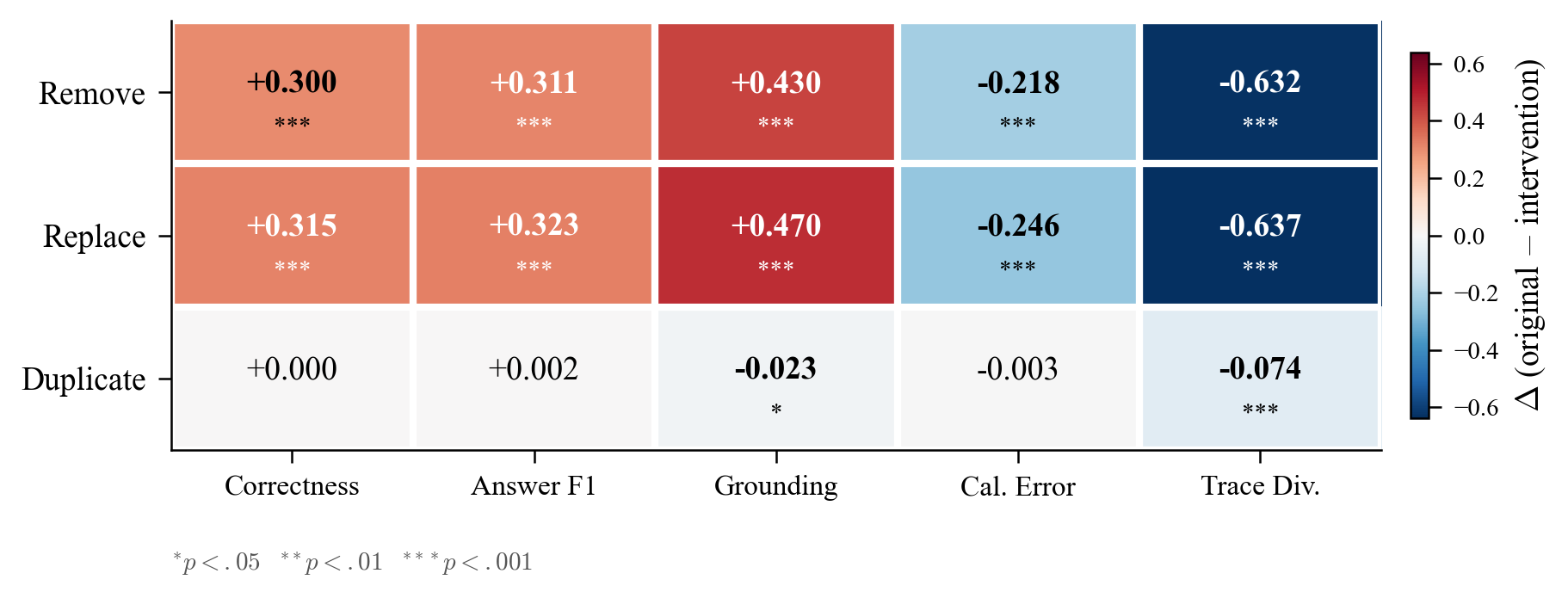}
\caption{\textbf{Paired bootstrap delta heatmap} for Qwen-3 8B on HotpotQA ($n=200$). Color encodes effect magnitude and direction (red = original outperforms; blue = intervention worsens trace/confidence). Significance: $^*p<.05$, $^{**}p<.01$, $^{***}p<.001$. \textsc{Remove} and \textsc{replace} show large, significant effects across all axes. \textsc{Duplicate} is near-zero on correctness but significant on grounding and trace divergence.}
\label{fig:heatmap}
\end{figure}

\noindent This duplicate effect is not an artifact of insertion position: a separate position-sensitivity check (\Cref{app:dup-position}) shows that front, after-original, and end placements all produce non-zero evidence-use divergence, though the magnitude is position-dependent. The full delta structure is visualized in the heatmap in \Cref{fig:heatmap}.

\FloatBarrier
\subsection{Zero-Retrieval Control}
\label{sec:zero-result}

\begin{table}[!htbp]
\centering
\caption{Zero-retrieval control on HotpotQA with Qwen-3 8B ($n = 200$). Retrieval is genuinely beneficial; intervention effects reflect degradation of meaningful evidence.}
\label{tab:zero}
\small
\begin{tabular}{@{}lcccc@{}}
\toprule
\textbf{Setting} & \textbf{Correct.}$\uparrow$ & \textbf{Ans. F1}$\uparrow$ & \textbf{Ground.}$\uparrow$ & \textbf{Conf. Err.}$\downarrow$ \\
\midrule
Original Retrieval & 0.580 & 0.629 & 0.823 & 0.430 \\
Zero Retrieval     & 0.220 & 0.270 & 0.000 & 0.676 \\
\bottomrule
\end{tabular}
\end{table}

\begin{figure}[!htbp]
\centering
\includegraphics[width=0.8\linewidth]{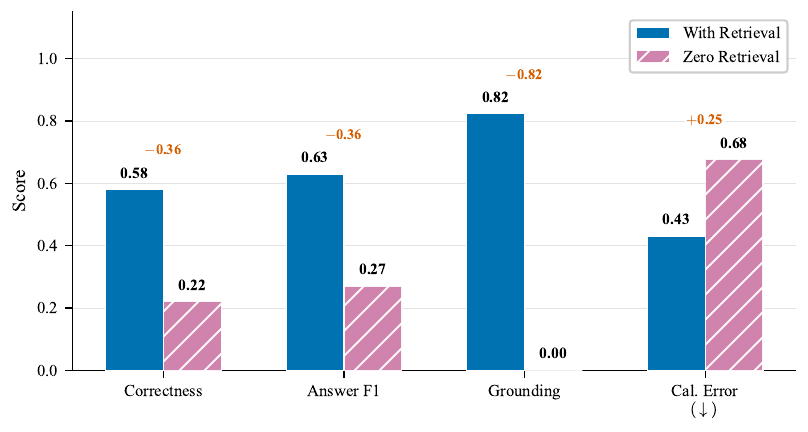}
\caption{\textbf{Zero-retrieval control} on HotpotQA with Qwen-3 8B ($n=200$). Retrieval substantially improves correctness ($+$0.36), F1 ($+$0.36), and grounding ($+$0.82). Red annotations show the drop from retrieval to zero-retrieval. Confidence error increases without retrieval.}
\label{fig:zero-retrieval}
\end{figure}

In a separate control experiment on 200 HotpotQA examples, the original retrieval condition achieves correctness 0.58 and answer F1 0.629, while the zero-retrieval baseline falls to correctness 0.22 and F1 0.270 (\Cref{tab:zero}). Grounding drops to 0.0 by construction since no evidence is provided. This confirms that retrieval is genuinely beneficial, and that the intervention effects in \Cref{sec:main-result,sec:stats-result} reflect degradation of meaningful evidence rather than arbitrary prompt instability (\Cref{fig:zero-retrieval}).

\FloatBarrier
\subsection{Cross-Dataset Replication: 2WikiMultihopQA (Q3)}
\label{sec:2wiki-result}

\begin{table}[!htbp]
\centering
\caption{\cuer{} results on 2WikiMultihopQA with Qwen-3 8B ($n = 100$). The same qualitative intervention ordering appears.}
\label{tab:2wiki}
\small
\begin{tabular}{@{}lccccc@{}}
\toprule
\textbf{Intervention} & \textbf{Correct.}$\uparrow$ & \textbf{Ans. F1}$\uparrow$ & \textbf{Ground.}$\uparrow$ & \textbf{Conf. Err.}$\downarrow$ & \textbf{Trace Div.}$\downarrow$ \\
\midrule
Original  & 0.540 & 0.538 & 0.818 & 0.399 & 0.000 \\
\midrule
Remove    & 0.390 & 0.388 & 0.465 & 0.502 & 0.594 \\
Replace   & 0.370 & 0.371 & 0.426 & 0.539 & 0.622 \\
Duplicate & \textbf{0.510} & \textbf{0.508} & \textbf{0.840} & \textbf{0.448} & \textbf{0.063} \\
\bottomrule
\end{tabular}
\end{table}

On 2WikiMultihopQA (\Cref{tab:2wiki}), the same qualitative intervention ordering appears. Original correctness is 0.540; removal reduces it to 0.390 ($-$0.150) and replacement to 0.370 ($-$0.170), while duplication is milder at 0.510 ($-$0.030). Trace divergence is 0.594 under removal and 0.622 under replacement, but only 0.063 under duplication. Grounding drops sharply under harmful interventions (from 0.818 to 0.465 and 0.426), reinforcing that the pattern generalizes beyond HotpotQA.

\finding{The intervention ordering (original $>$ duplicate $\gg$ remove $\geq$ replace) is consistent across both datasets, suggesting that these effects are not purely dataset-specific.}

\FloatBarrier
\subsection{Cross-Model Replication: GPT-5.2 (Q4)}
\label{sec:gpt-result}

\begin{table}[!htbp]
\centering
\caption{\cuer{} results on HotpotQA with GPT-5.2 ($n = 100$). The qualitative pattern persists; GPT-5.2 shows a higher baseline and relatively smaller answer-quality gaps but large trace divergence under harmful interventions.}
\label{tab:gpt}
\small
\begin{tabular}{@{}lccccc@{}}
\toprule
\textbf{Intervention} & \textbf{Correct.}$\uparrow$ & \textbf{Ans. F1}$\uparrow$ & \textbf{Ground.}$\uparrow$ & \textbf{Conf. Err.}$\downarrow$ & \textbf{Trace Div.}$\downarrow$ \\
\midrule
Original  & 0.690 & 0.746 & 0.878 & 0.309 & 0.000 \\
\midrule
Remove    & 0.480 & 0.530 & 0.575 & 0.408 & 0.525 \\
Replace   & 0.490 & 0.555 & 0.533 & 0.418 & 0.552 \\
Duplicate & \textbf{0.670} & \textbf{0.741} & \textbf{0.880} & \textbf{0.338} & \textbf{0.077} \\
\bottomrule
\end{tabular}
\end{table}

\begin{table}[!htbp]
\centering
\caption{Paired bootstrap deltas for GPT-5.2 on HotpotQA ($n = 100$; 5{,}000 resamples). With larger samples, GPT-5.2 shows significant correctness effects alongside grounding and trace effects. $^\dagger$Significant at $p < 0.05$.}
\label{tab:gpt-deltas}
\small
\begin{tabular}{@{}llccc@{}}
\toprule
\textbf{Comparison} & \textbf{Metric} & $\boldsymbol{\Delta}$ & \textbf{95\% CI} & $\boldsymbol{p}$ \\
\midrule
\multirow{5}{*}{Orig vs Remove}
  & Correctness & $+$0.210 & [0.120, 0.310] & $<$0.001$^\dagger$ \\
  & Answer F1   & $+$0.215 & [0.128, 0.307] & $<$0.001$^\dagger$ \\
  & Grounding   & $+$0.303 & [0.227, 0.382] & $<$0.001$^\dagger$ \\
  & Conf.\ Error & $-$0.099 & [$-$0.173, $-$0.027] & 0.009$^\dagger$ \\
  & Trace Div.  & $-$0.525 & [$-$0.569, $-$0.479] & $<$0.001$^\dagger$ \\
\midrule
\multirow{5}{*}{Orig vs Replace}
  & Correctness & $+$0.200 & [0.100, 0.300] & $<$0.001$^\dagger$ \\
  & Answer F1   & $+$0.191 & [0.096, 0.289] & $<$0.001$^\dagger$ \\
  & Grounding   & $+$0.345 & [0.267, 0.427] & $<$0.001$^\dagger$ \\
  & Conf.\ Error & $-$0.109 & [$-$0.181, $-$0.036] & 0.002$^\dagger$ \\
  & Trace Div.  & $-$0.552 & [$-$0.598, $-$0.506] & $<$0.001$^\dagger$ \\
\midrule
\multirow{5}{*}{Orig vs Duplicate}
  & Correctness & $+$0.020 & [$-$0.030, 0.070] & 0.520 \\
  & Answer F1   & $+$0.005 & [$-$0.035, 0.046] & 0.808 \\
  & Grounding   & $-$0.002 & [$-$0.032, 0.030] & 0.945 \\
  & Conf.\ Error & $-$0.029 & [$-$0.068, 0.005] & 0.105 \\
  & Trace Div.  & $-$0.077 & [$-$0.111, $-$0.047] & $<$0.001$^\dagger$ \\
\bottomrule
\end{tabular}
\end{table}

The qualitative structure remains intact with GPT-5.2 (\Cref{tab:gpt}). Original correctness is 0.690. Removal reduces correctness to 0.480 and replacement to 0.490, while duplication is milder at 0.670. Trace divergence shows a clear separation: 0.525 for remove, 0.552 for replace, and only 0.077 for duplicate.

With larger samples ($n = 100$), both correctness deltas are now statistically significant (\Cref{tab:gpt-deltas}): remove ($+$0.210, $p < 0.001$) and replace ($+$0.200, $p < 0.001$). Nonetheless, GPT-5.2 retains a higher baseline (0.690 vs.\ 0.585) and shows relatively smaller proportional drops than Qwen-3 8B. \cuer{} detects substantial grounding drops (remove: $+$0.303, $p < 0.001$; replace: $+$0.345, $p < 0.001$) and significant confidence-error worsening under both harmful interventions. The duplicate intervention remains answer-neutral ($+$0.020 correctness, $p = 0.520$), yet trace divergence is significant ($-$0.077, $p < 0.001$).

\finding{Stronger models show higher baselines and relatively smaller answer drops, but \cuer{} detects significant correctness, grounding, and trace-level disruption across all harmful interventions for both model families.}

\begin{figure}[!htbp]
\centering
\includegraphics[width=\linewidth]{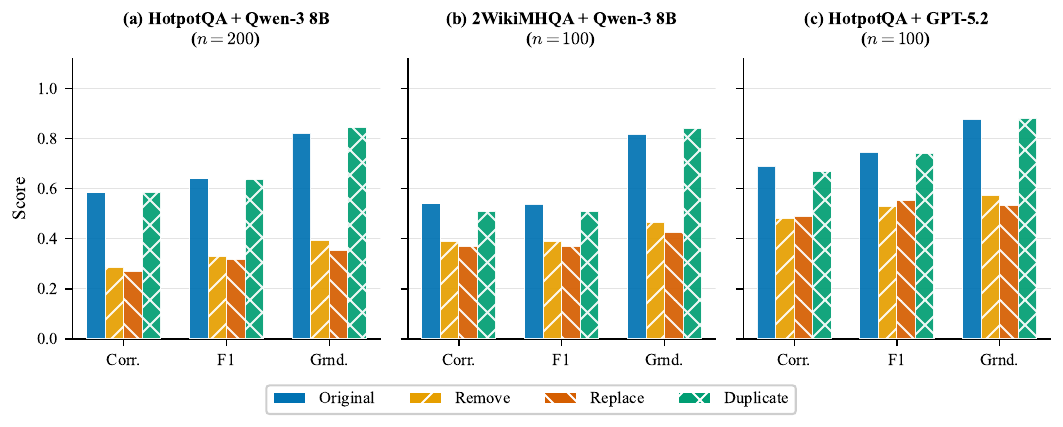}
\caption{\textbf{Cross-dataset and cross-model replication.} The intervention ordering (original $>$ duplicate $\gg$ remove $\geq$ replace) is consistent across HotpotQA + Qwen-3 8B ($n=200$), 2WikiMultihopQA + Qwen-3 8B ($n=100$), and HotpotQA + GPT-5.2 ($n=100$). All harmful intervention effects are statistically significant across configurations.}
\label{fig:cross-comparison}
\end{figure}

\noindent \Cref{fig:cross-comparison} visualizes the cross-dataset and cross-model replication side by side.

\FloatBarrier
\subsection{Replacement-Hardness Sweep}
\label{sec:sweep-result}

\begin{table}[!htbp]
\centering
\caption{Replacement-hardness sweep on HotpotQA with Qwen-3 8B ($n \approx 99$ per condition). All replacement strategies are consistently disruptive; correctness is identical across hardness levels.}
\label{tab:sweep}
\small
\begin{tabular}{@{}lccccc@{}}
\toprule
\textbf{Hardness} & \textbf{Correct.}$\uparrow$ & \textbf{Ans. F1}$\uparrow$ & \textbf{Ground.}$\uparrow$ & \textbf{Conf. Err.}$\downarrow$ & \textbf{Trace Div.}$\downarrow$ \\
\midrule
Easy   & 0.354 & 0.394 & 0.397 & 0.580 & 0.633 \\
Medium & 0.354 & 0.394 & 0.381 & 0.609 & 0.622 \\
Hard   & 0.354 & 0.416 & 0.434 & 0.585 & 0.616 \\
\bottomrule
\end{tabular}
\end{table}

\begin{figure}[!htbp]
\centering
\includegraphics[width=0.9\linewidth]{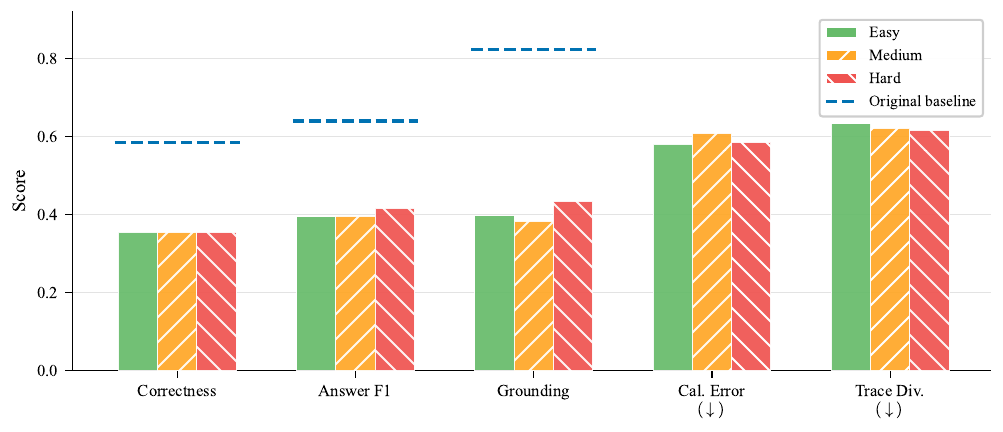}
\caption{\textbf{Replacement-hardness sweep} on HotpotQA with Qwen-3 8B ($n \approx 99$ per condition). Dashed blue lines indicate the original (unperturbed) baseline for the first three metrics. All three replacement strategies (easy, medium, hard) are consistently harmful, with identical correctness across hardness levels.}
\label{fig:sweep}
\end{figure}

\Cref{tab:sweep} shows results across three replacement strategies. All three are consistently harmful relative to the original baseline (correctness 0.585). All three hardness levels produce \emph{identical} correctness (0.354), though they differ on secondary metrics. Hard replacement (target-similar chunk) yields slightly higher F1 (0.416) and grounding (0.434) than easy and medium, suggesting that target-similar distractors may partially preserve some relevant context structure. The differences between hardness levels are small compared to the gap from the original baseline, and all three produce similarly large trace divergence (0.62-0.63). The main conclusion is that \textbf{evidence corruption is consistently disruptive regardless of replacement strategy} (\Cref{fig:sweep}). In other words, the current hardness sweep is less informative than the broader intervention-type comparison, suggesting that intervention type matters more than our present easy/medium/hard replacement taxonomy.

\FloatBarrier
\subsection{Two-Support Synergy Ablation (Q5)}
\label{sec:synergy-result}

To test whether \cuer{} can reveal multi-hop interaction effects beyond single-item necessity, we performed a two-support ablation on 51 HotpotQA examples where at least two gold support chunks were retrieved. For each example, we separately removed support~1, support~2, and both supports, then measured answer F1 drop relative to the original.

\begin{table}[!htbp]
\centering
\caption{Two-support synergy ablation on HotpotQA with Qwen-3 8B ($n = 51$ eligible examples). Joint removal causes a substantially larger F1 drop than either single removal. Synergy$_\text{max}$ denotes the excess joint drop over the worse single removal.}
\label{tab:synergy}
\small
\begin{tabular}{@{}lc@{}}
\toprule
\textbf{Statistic} & \textbf{Value} \\
\midrule
Mean F1 drop (support 1 only)       & 0.205 \\
Mean F1 drop (support 2 only)       & 0.186 \\
Mean F1 drop (both supports)        & 0.493 \\
\midrule
Mean synergy over max single drop   & 0.046 \\
\% examples with positive synergy   & 19.6\% \\
\% strong complementary (both single removals harmless, joint removal harmful) & 13.7\% \\
\midrule
$n$ eligible examples               & 51 \\
\bottomrule
\end{tabular}
\end{table}

Removing both supports caused a mean F1 drop of 0.493, compared with 0.205 and 0.186 for the individual removals (\Cref{tab:synergy}). In 19.6\% of eligible examples, the joint removal harmed performance more than the worse single removal, indicating non-additive interaction. Several examples exhibited a strong complementary pattern: neither single removal changed the answer, but removing both caused failure (13.7\% of cases). For instance, in a question requiring both the \emph{Animorphs} article and \emph{The Hork-Bajir Chronicles} article to identify the correct series, removing either support alone left the answer intact, but removing both caused the model to select a different (wrong) series entirely.

\finding{In multi-hop questions, evidence items can interact non-additively. Single-item interventions may understate the true dependence on retrieved evidence when supports are jointly necessary.}

\FloatBarrier
\subsection{Qualitative Case Studies}
\label{sec:cases}

We present representative case studies illustrating the evidence-role taxonomy (\Cref{tab:cases}).

\begin{table}[!htbp]
\centering
\caption{Representative qualitative case studies from HotpotQA + Qwen-3 8B.}
\label{tab:cases}
\small
\begin{tabular}{@{}p{2.0cm}p{3.8cm}cccc@{}}
\toprule
\textbf{Evidence Role} & \textbf{Question (abridged)} & \textbf{Orig.} & \textbf{Perturbed} & \textbf{$\Delta$Corr.} & \textbf{Trace Div.} \\
\midrule
Constructive & Brown State Fishing Lake \ldots population? & 9,984 $\checkmark$ & Unknown $\times$ & $+$1.0 & 0.98 \\
\addlinespace
Answer-preserving but trace-divergent & Japanese manga \ldots born what year? & 1970 $\times$ & 1968 $\times$ & 0.0 & 0.88 \\
\addlinespace
Redundant & Were Derrickson and Ed Wood same nationality? & no $\times$ & no $\times$ & 0.0 & 0.00 \\
\addlinespace
Conf.-distorting & Higher instrument ratio, Badly Drawn Boy or Wolf Alice? & BDB $\checkmark$ (0.9) & BDB $\checkmark$ (0.5) & 0.0 & 0.08 \\
\bottomrule
\end{tabular}
\end{table}

\paragraph{Constructive evidence.} For the question ``Brown State Fishing Lake is in a county that has a population of how many inhabitants?'' (gold: 9,984), the target chunk \emph{Brown County, Kansas} is critical. The original system correctly answers 9,984 with confidence 0.9. Removing this chunk causes the answer to collapse to ``Unknown'' with confidence 0.0 and trace divergence $\approx$1.0.

\paragraph{Trace-divergent, answer-preserving.} For a question about the birth year of a manga illustrator (gold: 1962), the original answer (1970) is incorrect. Under replacement, the answer shifts to 1968 (still incorrect), but confidence jumps from 0.5 to 0.9 and trace divergence reaches 0.88. The binary correctness label is unchanged in both cases, yet the system's trajectory diverges substantially. Answer-only evaluation would miss this entirely.

\paragraph{Duplicate-redundant.} For ``Were Scott Derrickson and Ed Wood of the same nationality?'' (gold: yes), the original answer is ``no'' (incorrect). Duplication produces the identical answer with identical confidence (0.9), grounding (0.5), and trace divergence = 0.0. A clear redundant case.

\paragraph{Confidence-distorting.} For a question about instrument-to-person ratios, the original answer is correct (Badly Drawn Boy, confidence 0.9). Duplication preserves the correct answer but drops confidence to 0.5, increasing confidence error from 0.1 to 0.5 while trace divergence remains low (0.08).

\FloatBarrier
\subsection{Summary of Findings}
\label{sec:summary}

Across datasets and models, four patterns are consistent:
\begin{enumerate}[leftmargin=*,itemsep=2pt]
  \item \textbf{\textsc{Remove} and \textsc{replace} are strongly harmful}, reducing correctness, grounding, and confidence alignment while causing large trace divergence.
  \item \textbf{\textsc{Duplicate} is often much milder on correctness}, but is not always neutral: it can still alter trace behavior and confidence error.
  \item \textbf{Trace-sensitive evaluation exposes evidence effects} that answer-only metrics alone do not cleanly separate.
  \item \textbf{Evidence items can interact non-additively} in multi-hop questions: joint removal causes degradation that exceeds either single removal, revealing complementary dependence.
\end{enumerate}

\section{Limitations}
\label{sec:limitations}

\cuer{} is intentionally narrow in scope, and several limitations follow from that design:

\begin{itemize}[leftmargin=*,itemsep=3pt]
  \item \textbf{Interventional sensitivity, not strong causality.} Our perturbations modify the input prompt, which changes prompt length, context distribution, and attention allocation simultaneously. The measured effects are best understood as \emph{interventional sensitivity under prompt-level evidence perturbation}, not as causal contribution in the strongest counterfactual sense. We use ``intervention-based utility'' as operational shorthand throughout, but readers should interpret it accordingly.

  \item \textbf{Observable traces only.} \cuer{} evaluates observable traces, not hidden internal reasoning. The framework measures how evidence perturbations change logged behavior and downstream utility, but does not claim to recover the model's true internal mechanism~\citep{lanham2023measuring}.

  \item \textbf{Shallow traces.} Our current experiments use shallow single-shot traces rather than full agentic multi-step workflows. Extending \cuer{} to deeper agentic settings~\citep{yao2023react,liu2024agentbench} is an important direction.

  \item \textbf{Proxy-based grounding.} The grounding score is a coarse proxy based on title-level overlap between model-used chunk identifiers and gold support titles. It does not verify whether the model used the correct information within a chunk, and gold titles may not exhaustively enumerate all useful evidence. More fine-grained grounding metrics (e.g., sentence-level attribution) would strengthen the analysis.

  \item \textbf{Self-reported confidence.} The confidence error metric relies on the model's self-reported confidence, which is known to be noisy and poorly calibrated in many LLMs~\citep{kadavath2022language}. Our per-instance absolute error is a lightweight proxy, not a distributional calibration analysis.

  \item \textbf{Limited intervention set.} The current framework uses only \textsc{remove}, \textsc{replace}, and \textsc{duplicate}. Additional perturbation types (\eg, paraphrase, partial corruption, contradiction injection) could further enrich the taxonomy.

  \item \textbf{Limited multi-item interventions.} Our main experiments perturb one evidence item at a time. The two-support ablation (\Cref{sec:synergy-result}) provides initial evidence of non-additive interaction, but systematic combinatorial interventions across all evidence pairs remain future work.

  \item \textbf{Simple retrieval backbone.} We use BM25~\citep{robertson2009probabilistic} with top-$k$ passage selection rather than a state-of-the-art dense retriever~\citep{karpukhin2020dense}. This is intentional for reproducibility but limits generalization claims.

  \item \textbf{Empirical scale.} While we replicate the main pattern across two datasets and two model families with 100--200 examples per condition, larger-scale validation across additional domains and model families would further strengthen generalization claims.

  \item \textbf{Heuristic target selection.} Target evidence items are selected using a priority heuristic that favors model-used and support-aligned chunks. Different target-selection policies could yield different intervention profiles.

  \item \textbf{Operational taxonomy.} The evidence-role taxonomy should be interpreted as an operational diagnostic framework, not an ontology of model cognition.
\end{itemize}

\section{Conclusion}
\label{sec:conclusion}

We introduced \cuer{}, a lightweight intervention-based framework that asks what operational utility a retrieved evidence item contributed once a RAG system acted on it. By combining observable retrieval-use traces, three evidence perturbation operators, and multi-axis utility analysis, the framework surfaces behavioral effects that answer-only metrics miss.

Across HotpotQA and 2WikiMultihopQA (100--200 examples per condition), Qwen-3 8B and GPT-5.2, the central finding is consistent: \textsc{remove} and \textsc{replace} interventions sharply reduce correctness and grounding while inducing large trace divergence, whereas \textsc{duplicate} interventions are milder on answer quality yet still measurably alter confidence error and trace behavior. A two-support ablation further shows that evidence items can interact non-additively in multi-hop questions, with joint removal causing far greater degradation than either single removal. A zero-retrieval control confirms that these effects reflect degradation of meaningful evidence, not arbitrary prompt instability.

We do not claim that \cuer{} recovers hidden model cognition. \cuer{} should be read less as a claim about model cognition and more as a \textbf{practical diagnostic tool} for identifying which retrieved items are useful, redundant, or behaviorally destabilizing. The operational taxonomy it produces is of a convenient forensic value, and not an ontology.

The current instantiation is limited to shallow single-shot traces with proxy-based grounding and self-reported confidence. Future work should extend \cuer{} to richer agentic settings with deeper traces, stronger grounding metrics, sentence-level interventions, and dense retrievers.

\clearpage
\bibliographystyle{unsrtnat}
\bibliography{references}

\appendix

\section{Bootstrap Confidence Intervals}
\label{app:ci}

\begin{table}[H]
\centering
\caption{Intervention means with 95\% bootstrap CIs for Qwen-3 8B on HotpotQA ($n = 200$).}
\label{tab:ci-qwen}
\small
\begin{tabular}{@{}lcccc@{}}
\toprule
\textbf{Intervention} & \textbf{Correctness} & \textbf{Answer F1} & \textbf{Grounding} & \textbf{Conf. Error} \\
\midrule
Original  & 0.585 [0.52, 0.66] & 0.640 [0.578, 0.701] & 0.823 [0.77, 0.87] & 0.422 [0.37, 0.48] \\
Remove    & 0.285 [0.23, 0.35] & 0.329 [0.271, 0.393] & 0.392 [0.33, 0.46] & 0.639 [0.59, 0.69] \\
Replace   & 0.270 [0.21, 0.34] & 0.318 [0.259, 0.381] & 0.353 [0.29, 0.42] & 0.667 [0.61, 0.72] \\
Duplicate & 0.585 [0.52, 0.65] & 0.639 [0.577, 0.699] & 0.845 [0.80, 0.89] & 0.424 [0.37, 0.48] \\
\bottomrule
\end{tabular}
\end{table}

\begin{table}[H]
\centering
\caption{Intervention means with 95\% bootstrap CIs for GPT-5.2 on HotpotQA ($n = 100$).}
\label{tab:ci-gpt}
\small
\begin{tabular}{@{}lcccc@{}}
\toprule
\textbf{Intervention} & \textbf{Correctness} & \textbf{Answer F1} & \textbf{Grounding} & \textbf{Conf. Error} \\
\midrule
Original  & 0.690 [0.60, 0.78] & 0.746 [0.668, 0.817] & 0.878 [0.82, 0.93] & 0.309 [0.25, 0.37] \\
Remove    & 0.480 [0.38, 0.58] & 0.530 [0.439, 0.620] & 0.575 [0.49, 0.67] & 0.408 [0.36, 0.46] \\
Replace   & 0.490 [0.39, 0.59] & 0.555 [0.466, 0.642] & 0.533 [0.44, 0.63] & 0.418 [0.37, 0.47] \\
Duplicate & 0.670 [0.58, 0.76] & 0.741 [0.664, 0.814] & 0.880 [0.83, 0.93] & 0.338 [0.27, 0.41] \\
\bottomrule
\end{tabular}
\end{table}

\section{Metric Definitions}
\label{app:metrics}

\paragraph{Soft Correctness.} Given predicted answer $y$ and gold answer $y^*$, soft correctness is 1 if any of the following hold: (a)~$\textsc{norm}(y) = \textsc{norm}(y^*)$, (b)~both canonicalize to the same yes/no value, (c)~both are numerically equivalent, or (d)~token-level $\text{F1}(y, y^*) \geq 0.8$.

\paragraph{Answer F1.} Token-level F1 computed as $\frac{2PR}{P+R}$ where $P = \frac{|\text{common}|}{|\text{pred\_tokens}|}$ and $R = \frac{|\text{common}|}{|\text{gold\_tokens}|}$ after normalization.

\paragraph{Grounding Score.} Defined in \Cref{eq:grounding}.

\paragraph{Confidence Error.} $\text{CE} = |c - \mathbb{1}[\text{is\_correct}]|$, where $c \in [0,1]$ is the model's self-reported confidence~\citep{guo2017calibration,kadavath2022language}, clamped to the unit interval. This is an instance-level absolute error, not a distributional calibration metric.

\paragraph{Trace Divergence.} Defined in \Cref{eq:div-proxy}. For the Jaccard component, $d_{\text{Jaccard}}(E, E') = 1 - \frac{|E \cap E'|}{|E \cup E'|}$, where $E$ and $E'$ are the sets of used chunk identifiers in the original and perturbed runs, respectively.

\section{Duplicate Position Sensitivity}
\label{app:dup-position}

A reviewer might ask whether the duplicate result is an artifact of always appending the copied chunk at the end. To check this, we tested three insertion positions on 60 HotpotQA examples with Qwen-3 8B: front (before all chunks), after-original (immediately after the original copy), and end (after all chunks, the default).

\begin{table}[H]
\centering
\caption{Duplicate position sensitivity on HotpotQA with Qwen-3 8B ($n = 60$). All placements remain largely answer-preserving but produce non-zero trace divergence. Front and after-original produce larger shifts than end.}
\label{tab:dup-position}
\small
\begin{tabular}{@{}lccccc@{}}
\toprule
\textbf{Condition} & \textbf{Correct.}$\uparrow$ & \textbf{Ans.\ F1}$\uparrow$ & \textbf{Ground.}$\uparrow$ & \textbf{Trace Div.} & \textbf{Evid.\ Div.} \\
\midrule
Original            & 0.583 & 0.619 & 0.825 & 0.000 & 0.000 \\
\midrule
Duplicate (front)   & 0.567 & 0.620 & 0.808 & 0.086 & 0.094 \\
Duplicate (after)   & 0.583 & 0.630 & 0.825 & 0.076 & 0.083 \\
Duplicate (end)     & 0.567 & 0.613 & 0.808 & 0.045 & 0.050 \\
\bottomrule
\end{tabular}
\end{table}

Duplication remained largely answer-preserving across all three placements, but consistently induced non-zero evidence-use divergence (\Cref{tab:dup-position}). Front and after-original duplication produced larger trace shifts than end duplication, suggesting that the duplicate effect is partly mediated by positional salience rather than redundancy alone. Answer-level changes were small in all cases (correctness changed in at most 5\% of examples for front, 6.7\% for after-original, 1.7\% for end), confirming that the main duplicate finding is not an artifact of one insertion scheme.

\section{Prompt Template}
\label{app:prompt}

All models receive the following structured prompt (adapted for each condition):

\begin{verbatim}
You are a QA system. Answer the question using ONLY
the provided context.

Return a JSON object with exactly these keys:
  "answer": your short answer,
  "confidence": float between 0 and 1,
  "used_chunk_ids": list of chunk IDs you relied on,
  "brief_reason": one sentence explanation.

Context:
[ctx_0] Title: ... Content: ...
[ctx_1] Title: ... Content: ...
...

Question: {question}

Respond with valid JSON only.
\end{verbatim}

For the zero-retrieval control, the context section is replaced with: ``Context: (No retrieved documents available. Answer from your own knowledge.)''

\section{Reproducibility Details}
\label{app:reproducibility}

\Cref{tab:reproducibility} summarizes the key experimental parameters for reproducibility.

\begin{table}[H]
\centering
\caption{Reproducibility details for all experiments.}
\label{tab:reproducibility}
\small
\begin{tabular}{@{}ll@{}}
\toprule
\textbf{Parameter} & \textbf{Value} \\
\midrule
Qwen-3 8B model & \texttt{qwen3:8b} via Ollama (local) \\
GPT-5.2 model & \texttt{gpt-5.2} (API) \\
Decoding temperature & 0 (both models) \\
Retrieval & BM25, top $k = 5$ \\
HotpotQA split & Distractor setting (dev set) \\
2WikiMultihopQA split & Dev set \\
Bootstrap resamples & 5{,}000 (seed = 42) \\
Confidence interval & 95\%, percentile method \\
Experiment period & January--March 2026 \\
Prompt version & See \Cref{app:prompt} \\
\bottomrule
\end{tabular}
\end{table}

\end{document}